\documentclass[12pt,twoside,a4paper]{article}

\usepackage{amsmath}
\usepackage{a4wide}
\usepackage{graphicx}
\usepackage{dsfont}
\usepackage{amsthm}
\usepackage{amssymb}

\newcommand{\rem}[1]{}
\newcommand{\todo}[1]{}

\newcommand{\ui}{\mathrm{i}}

\newcommand{\vp}{{\mathbf{p}}}
\newcommand{\vq}{\mathbf{q}}
\newcommand{\vz}{\mathbf{z}}

\newcommand{\vJ}{{\mathbf{J}}}
\newcommand{\mM}{{\text{M}}}
\newcommand{\mR}{{\text{R}}}
\newcommand{\R}{\mathds{R}}

\newcommand{\dof}{{f}}

\newcommand{\Veff}{{V_{\text{eff}}}}

\title{Phase space structures governing reaction dynamics in rotating molecules}
\author{\"{U}nver \c{C}ift\c{c}i\thanks{uciftci@nku.edu.tr} $^1$ and Holger Waalkens\thanks{h.waalkens@rug.nl} $^2$ \\[2ex]
$^1$Department of Mathematics\\ Nam\i k Kemal University\\ 59030\\ Tekirda\u{g}, Turkey\\[1.5ex]
$^2$Johann Bernoulli Institute for Mathematics and Computer Science\\ University of Groningen\\ PO Box 407\\ 
9700 AK Groningen, The Netherlands}

\begin{document}
\maketitle

\abstract{
Recently the phase space structures governing reaction dynamics in Hamiltonian systems have been identified and algorithms for their explicit construction have been developed. 
These phase space structures are induced by  saddle type equilibrium points which are characteristic for reaction type dynamics. 
Their construction is based on a Poincar{\'e}-Birkhoff normal form. 
Using  tools from the geometric theory of Hamiltonian systems and their reduction 
we show in this paper how the construction of these phase space structures can be generalized to the case of the relative equilibria of a rotational symmetry reduced $N$-body system.
As rotations almost always play an important role in the reaction dynamics of molecules
the approach presented in this paper is of great relevance for applications. 
}


\vspace*{1cm}


\section{Introduction}

From the perspective of dynamical systems theory a dynamical system shows reaction type dynamics if its  phase space 
possesses a bottleneck type geometry. The system spends a long time in one phase space region before it finds its way through a bottleneck to another phase space regions.
In the terminology of chemistry these phase space regions correspond to `reactants' and `products', respectively, and the bottleneck connecting the two is called a `transition state'.
For Hamiltonian systems, the phase space bottlenecks result from equilibrium points of a certain stability type, namely equilibria for which the matrix associated with the linearization 
has one pair real eigenvalues of opposite sign and $\dof-1$ complex conjugate pairs of imaginary eigenvalues, where $\dof$ is the number of degrees of freedom. Such an equilibrium is called a saddle$\times$center$\times\cdots \times$center, and we will refer to them as `saddle' for short.
Near a saddle the energy surface (locally) bifurcates from an energy surface with two disjoint components (the reactants and products) for energies below the energy of the saddle 
to an energy surface consisting of a single connected component for energies above the energy of the saddle. The single component energy surface has a wide-narrow-wide geometry, and the system has to pass or `react' through this bottleneck in order to evolve from reactants to products or vice versa. The most widely used approach to compute reaction rates in chemistry is to place a dividing surface in the bottleneck region and compute the rate from the flux through this dividing surface. 
This forms the basis of \emph{Transition State Theory} that was developed by Eyring, Polanyi and Wigner in the 1930's \cite{Eyring35,Wigner38}.
For this approach to be useful the dividing surface needs to have the property that it is crossed exactly once by all reactive trajectories (i.e., trajectories moving from reactants to products or vice versa) and not crossed at all by the non-reactive trajectories (i.e., trajectories staying on the reactants or products side). The construction of a dividing surface which solves this so called recrossing problem has posed a major difficulty in the development of transition state theory. In the 1970's Pechukas,  Pollak and others \cite{PechukasMcLafferty73,PechukasPollak78} showed that for systems with two degrees of freedom, a recrossing free dividing surface can be constructed from a periodic orbit (the Lyapunov orbit associated with the saddle). The longstanding problem of how to generalize this idea for systems with more than two degrees of freedom has been solved only recently  \cite{Wigginsetal01,Uzeretal02}, see also \cite{MacKay1990,KomaBerry1999}. Here the role of the Lyapunov orbit  is taken over by a \emph{normally hyperbolic invariant manifold} (NHIM) \cite{Wiggins90,Wiggins94} which forms the anchor for the construction of a dividing surface in the general case. The NHIM also gives a precise meaning of the transition state. It can be viewed as the energy surface of an invariant subsystem with $n-1$ degrees of freedom, which as an unstable `super molecule'  \cite{Pechukas76} is located between reactants and products. The NHIM is not only of central significance for the construction of a recrossing free dividing surface but also gives detailed information of the geometry and the mechanism of reactions. This arises from the fact that the NHIM has stable and unstable manifolds which are of one dimension less than the energy surface. They form the separatrices which separate the reactive trajectories and non-reactive trajectories in the energy surface. In fact they form tubes which snake through the phase space and for a phase space point in the region of reactants to be reactive (i.e., for a point leading to a reactive trajectory if it is taken as an initial condition for Hamilton' s equations) it has to be contained in a certain volume that is enclosed by the stable manifold of the NHIM. Similarly only points in the region of products can be reached by reactive trajectories emanating from the reactants if they are contained in a certain volume enclose by the unstable manifold of the NHIM. This information is crucial not only for the study of state specific reactivities but also for the control of reactions. 
Most importantly for applications, 
the dividing surface, the NHIM and the local pieces of its stable and unstable manifolds can be explicitly constructed from a  Poincar{\'e}-Birkhoff normal form expansion about the saddle equilibrium point \cite{Uzeretal02,Waalkensetal08}.  
The normal form gives a symplectic transformation to a new set of phase space coordinates in terms of which one can give simple formulas for the phase space structures. The inverse of the normal form transformation then allows one to construct the phase space structures in the original (`physical') coordinates. 

In this paper we address the question of how the phase space structures can be constructed for the relative equilibria of a rotational symmetry reduced $N$-body system. 
Here we have in mind that the $N$ bodies represent the $N$ atoms that constitute a given molecule (described in the  Born-Oppenheimer approximation) but the theory presented also works for other $N$-body systems like  celestial $N$-body systems.
For molecular  reactions,  the excitation of rotational degrees of freedom almost always plays an important role. 
The phase space bottleneck are then no longer associated with equilibrium points but with larger sets of phase space points which however become again equilibria (so called \emph{relative equilbibria}) if considered in the rotational symmetry reduced system. 
In this context the reduction is important for both conceptual and computational reasons. 
We are particularly interested in a reduction which facilitates the Poincar{\'e}-Birkhoff normal construction of the phase space structures around a relative equilibrium in the same fashion as in the case of a usual saddle equilibrium. 
The theory of reduction for $N$-body systems is well developed \cite{Iwai87, JellinekLi89, Marsden92, KozinPavlichenko96, LittlejohnReinsch97, IwaiTachibana99}.  However, although the structure of the reduced space is well-known, the explicit choice  of  suitable coordinates for the reduced space remains as a challenging problem \cite{RobertsScmahStoica06}. If the angular momentum is zero, the reduced space is symplectomorphic to the tangent bundle of the \emph{shape} or \emph{internal space} (see Sec.~\ref{sec:Nbodyreduction}). For non-vanishing angular momentum however,  this is no longer the case  \cite{Iwai87}. This makes it difficult to apply classical techniques such as Poincar{\'e}-Birkhoff normal  form \cite{MeyerHall}, and that point is the motivation of our paper (see also \cite{Pringle73, Uzer90}). There exist a huge literature on the analysis of reduced $N$-body systems \cite{ChurchillKummerRod83, HarterPatterson84,ElipeFerre94, Hanssman95, MillerWales96, Sadovskii01, RobertsWulffLamb02, Yurtsever02, KozinRoberts03, WiesenfeldFaureJohann03, VeraVigueras, KozinSadovskiiZhilinskii05, SchmahStoica06, Gurfiletal07,  PatricRobertsWulff08, ArangoEzra08, YanaoKoonMarsden09, ChierchiaPinzari11}, and the theory is well-developed. Nevertheless a more canonical way of expressing the equations of motion seems to be useful, and we hope that the theory presented in this paper will also be of interest for studies of $N$-body systems other than the reaction dynamics studied here. 

In our approach we mainly follow the account given by Littlejohn and Reinsch \cite{LittlejohnReinsch97}. Their reduction procedure can be viewed as a generalization of the free rigid body reduction scheme. Namely, the Hamiltonian is written as the sum of rotational kinetic energy, vibrational kinetic energy and the potential by passing to the body coordinates. The only remaining coordinates are the body angular momentum, the shape or internal coordinates and the conjugate momenta of the  internal  coordinates. 
Using the body angular momentum the equations of motion have a non-canonical form which is not so suitable for the local analysis near equilibrium points \cite{KozinRobertsTennyson99, KozinRobertsTennyson00, LittlejohnMitchell03}.
Making use of the fact that the reduced space is locally a product of the constant body angular momentum sphere and the cotangent bundle of the internal space we write the equations of motion in a Hamiltonian form in this paper. We do this by choosing some suitable coordinates on the angular momentum sphere. 

This paper is organized as follows. In Sections~\ref{sec:phasespacestruc} and \ref{sec:Nbodyreduction}
we review the phase space structures governing reaction dynamics across saddle 
equilibrium points and  the reduction of the rotational symmetry of $N$-body systems, respectively. The main result of this paper is contained in Sec.~\ref{sec:canoncoord} where we introduce a canonical 
coordinate system on the symmetry reduced space which facilitates Poincar{\'e}-Birkhoff normal form computations near equilibrium points.
The theory is illustrated for the limiting case of a rigid body and the triatomic example  in Sec.~\ref{sec:examples}.
Conclusions and an outlook are given in Sec.\ref{sec:conclusions}.

\section{Phase space structures governing reactions across saddles}
\label{sec:phasespacestruc}

In this section we discuss the phase space structures governing reaction dynamics near saddle equilibrium points. Before studying the case of a general (nonlinear) Hamiltonian vector field it is useful to first consider the linear case. 
For the details we refer to  \cite{Uzeretal02, Waalkensetal08}.

\subsection{The linear case}
\label{sec:phasespacestruc_linear}

Consider a linear Hamiltonian vector field with a saddle on the phase space $\R^f\times\R^f$ where $f\ge 2$ (we comment on the case $f=1$ at the end of this section).  
Consider the quadratic Hamiltonian function
\begin{equation}\label{eq:Hquadratic}
H_2(\vq,\vp)= \frac{\lambda}{2} (p_1^2 - q_1^2) + \sum_{k=2}^f  \frac{\omega_k}{2} (p_k^2 + q_k^2) \,.
\end{equation}
The corresponding linear Hamiltonian vector field has a saddle equilibrium point at the origin, i.e. the matrix associated with the linear vector field has the pair of real eigenvalues $\pm \lambda$ and $f-1$ pairs of complex conjugate imaginary eigenvalues $\pm \ui \omega_k$, $k=2,\ldots,f$. We define the constants of motion
\begin{equation}\label{eq:def_constants_motion}
{\cal I}_1 = p_1^2 - q_1^2\,, \quad   I_k= p_k^2 + q_k^2 \,,\quad k=2,\ldots,\dof\,,
 \end{equation}
which up to positive prefactors  agree with the energies in the individual degrees of freedoms.

Consider a fixed energy $E>0$, where $0$ is the energy of the saddle. Rewriting the energy equation 
$H_2(\vq,\vp)=E$ in the form
\begin{equation}\label{eq:family_spheres}
\frac{\lambda}{2} p_1^2  + \sum_{k=2}^\dof  \frac{\omega_k}{2} (p_k^2 + q_k^2) = E +  \frac{\lambda}{2}  q_1^2
\end{equation}
one sees that each fixed $q_1\in \R$ defines a topological $(2\dof-2)$-dimensional sphere. The energy surface
\begin{equation}
\Sigma_E = \{ (\vq,\vp) \in \R^{2\dof} \,:\,  H_2(\vq,\vp)=E \}
\end{equation}
thus has the topology of a spherical cylinder $\R\times S^{2\dof-2}$. The `radius' of the family of $(2f-2)$-dimensional sphere \eqref{eq:family_spheres} becomes smallest for $q_1=0$, and this in fact can be used 
to define a recrossing free dividing surface (see the introduction). Setting
 $q_1=0$ on the energy surface gives the $(2\dof -2)$-dimensional sphere
\begin{equation}\label{eq:def_DS_linear}
S_{\text{DS}}^{2\dof -2}  = \{ (\vq,\vp) \in \R^{2\dof } \,:\,  H_2(\vq,\vp) = E \,, q_1=0 \}\,.
\end{equation}
The dividing surface $S_{\text{DS}}^{2\dof -2} $ divides the energy surface into the two components which have $q_1<0$ (the `reactants') and  $q_1>0$ (the `products'), respectively, and
as $\dot{q}_1= \partial H_2/\partial p_1=\lambda p_1\ne 0$ for $p_1\ne 0$  the dividing surface 
is everywhere transverse to the Hamiltonian flow except for the submanifold where $q_1=p_1=0$. For $q_1=p_1=0$, the energy equation \eqref{eq:family_spheres}  reduces to $\sum_{k=2}^\dof  \frac{\omega_k}{2} (p_k^2 + q_k^2) = E$. The submanifold thus is a  $(2\dof-3)$-dimensional sphere which we denote by
\begin{equation}\label{eq:NHIM_linear}
S_{\text{NHIM}}^{2\dof-3}  = \{ (\vq,\vp) \in \R^{2\dof} \,:\,  
H_2(\vq,\vp) = E \,, q_1=p_1=0  \}\,.
\end{equation}
This is a so called \emph{normally hyperbolic invariant manifold} \cite{Wiggins90,Wiggins94} (NHIM for short), i.e. $S_{\text{NHIM}}^{2\dof-3}$ is invariant (since $q_1=p_1=0$ implies $\dot{q}_1=\dot{p}_1=0$) and the contraction and expansion rates for motions on $S_{\text{NHIM}}^{2\dof-3}$ are dominated by those components related to directions transverse to $S_{\text{NHIM}}^{2\dof -3}$.  The NHIM \eqref{eq:NHIM_linear} can be considered to form the equator of the dividing surface \eqref{eq:def_DS_linear} in the sense that it divides into two hemispheres which topologically are $(2\dof-2)$-dimensional balls. All forward reactive trajectories (i.e. trajectories moving from reactants to products) cross one of these hemispheres, and all backward reactive trajectories (i.e. trajectories moving from products to reactants) cross the other of these hemispheres. 
Note that a trajectory is reactive only if it has ${\cal I}_1>0$ (i.e. if it has sufficient energy in the first degree of freedom). Trajectories with  ${\cal I}_1<0$ are nonreactive, i.e. they stay on the side of reactants or on the side of products.

Due to its normal hyperbolicity the NHIM
has stable and unstable manifolds which are given by setting
$p_1=-q_1$ resp.  $p_1=q_1$ on the energy surface. Each of them have two branches. We denote the branches of the stable manifold by
\begin{equation}
\begin{split}
W^{\text{s}}_{\text{NHIM;r}} &=  \{ (\vq,\vp) \in \R^{2\dof}   \,:\, H_2(\vq,\vp) = E,\, p_1=-q_1>0 \}\,, \\
W^{\text{s}}_{\text{NHIM;p}} &= \{ (\vq,\vp) \in \R^{2\dof} \,:\, H_2(\vq,\vp) = E,\, p_1=-q_1<0 \} \,,
\end{split}
\end{equation}
where $W^{\text{s}}_{\text{r}}$ is located on the reactants side of the dividing surface and $W^{\text{s}}_{\text{p}}$ is located on the product sides of the dividing surface.
Similarly the unstable manifold has the two branches
\begin{equation}
\begin{split}
W^{\text{u}}_{\text{NHIM;r}} &=  \{ (\vq,\vp)  \in \R^{2\dof} \,:\, H_2(\vq,\vp) = E,\, p_1=q_1<0 \} \,, \\
W^{\text{u}}_{\text{NHIM;p}} &= \{ (\vq,\vp)  \in \R^{2\dof} \,:\, H_2(\vq,\vp) = E,\, p_1=q_1>0 \} \,.
\end{split}
\end{equation}
The stable and unstable manifolds have the topology of spherical cylinders $\R\times S^{2\dof-3}$. As they are of co-dimension 1 in the energy surface they can act as separatrices, dividing the energy surface into different components.
In fact  on the stable and unstable manifolds  ${\cal I}_1$ is equal to zero. They thus lie `between' the non-reactive (${\cal I}_1<0$) and reactive (${\cal I}_1$>0) trajectories. More precisely the
forward reactive cylinder 
\begin{equation}
\text{C}_{\text{f}} =  W^{\text{s}}_{\text{NHIM;r}} \cup W^{\text{u}}_{\text{NHIM;p}}
\end{equation}
encloses all forward reactive trajectories in the energy surface and separates them from all other trajectories. Similarly the
backward reactive cylinder 
\begin{equation}
\text{C}_{\text{b}} =  W^{\text{s}}_{\text{NHIM;p}} \cup W^{\text{u}}_{\text{NHIM;r}}
\end{equation}
encloses all backward reactive trajectories in the energy surface and separates them from all other trajectories.  In this sense $C_{\text{f}} $ and $C_{\text{b}}$ form the phase space conduits of forward and backward reactions, respectively.

Moreover, it is useful to define the lines
\begin{equation}
\begin{split}
\text{DRP}_{\text{f}} &=  \{ (\vq,\vp)  \in \R^{2\dof} \,:\, H_2(\vq,\vp) = E,\, p_1>0, \, p_k=q_k=0, \, k=2,\ldots,\dof \}\,, \\
\text{DRP}_{\text{b}} &=  \{ (\vq,\vp)  \in \R^{2\dof} \,:\, H_2(\vq,\vp) = E,\, p_1<0, \, p_k=q_k=0, \, k=2,\ldots,\dof \}\,.
\end{split}
\end{equation}
These lines form the centerlines of the volumes enclosed by the forward and backward reactive cylinders $\text{C}_{\text{f}}$ and $\text{C}_{\text{b}}$, respectively, and are therefore referred to as
the forward and backward dynamical reaction paths, respectively. All forward reactive trajectories spiral about $\text{DRP}_{\text{f}}$, and all backward reactive trajectories spiral about $\text{DRB}_{\text{b}}$.
Note that in the limit $E\to 0^+ $ the forward and backward reactive cylinders $\text{C}_{\text{f}}$ and $\text{C}_{\text{b}}$ shrink to the
forward and backward dynamical reaction paths, which in turn become the one-dimensional stable and unstable manifolds of the saddle equilibrium point in this limit. 
More precisely, if 
\begin{equation}
\begin{split}
W^{\text{s}}_{\text{saddle;r}} &=  \{ (\vq,\vp)  \in \R^{2\dof} \,:\, H_2(\vq,\vp) = 0,\, p_1=-q_1>0, \, p_k=q_k=0, \, k=2,\ldots,\dof \} \,,\\
W^{\text{s}}_{\text{saddle;p}} &=  \{ (\vq,\vp)  \in \R^{2\dof} \,:\, H_2(\vq,\vp) = 0,\, p_1=-q_1<0, \, p_k=q_k=0, \, k=2,\ldots,\dof \} 
\end{split}
\end{equation}
denote the reactants and product branches of the stable manifold of the saddle, and
\begin{equation}
\begin{split}
W^{\text{u}}_{\text{saddle;r}} &=  \{ (\vq,\vp)  \in \R^{2\dof}  \,:\, H_2(\vq,\vp) = 0,\, p_1=q_1<0, \, p_k=q_k=0, \, k=2,\ldots,\dof \} \,, \\
W^{\text{u}}_{\text{saddle;p}} &=  \{ (\vq,\vp)  \in \R^{2\dof} \,:\, H_2(\vq,\vp) = 0,\, p_1=q_1>0, \, p_k=q_k=0, \, k=2,\ldots,\dof \} 
\end{split}
\end{equation}
denote the reactants and product branches of the unstable manifold of the saddle then
$\text{DRP}_{\text{f}}\to W^{\text{s}}_{\text{saddle;r}} \cup  W^{\text{u}}_{\text{saddle;p}}$ and $\text{DRP}_{\text{b}}\to W^{\text{s}}_{\text{saddle;p}} \cup  W^{\text{u}}_{\text{saddle;r}}$ for 
$E\to 0^+ $.
 
For systems with $\dof=1$ degree of freedom with a saddle, one can still separate the phase space (or energy surface) in a reactants and a products region in a similar manner as described above.
However, most of the phase space structures defined above make no sense for $f=1$. The case of one degree of freedom is special 
since in this case the trajectories are given by the level sets of the Hamiltonian. The question of whether a trajectory is reactive or not is thus completely determined by the (total) energy of the trajectory.
 
\subsection{The general (nonlinear) case}
\label{sec:phasespacestruc_nonlinear}

For the general nonlinear case, consider a Hamiltonian function $H$  which has an equilibrium point (a `saddle') at which 
the corresponding linearized vector field has the same disposition of eigenvalues as in Sec.~\ref{sec:phasespacestruc_linear}.
In the neighbourhood of the saddle the dynamics is thus similar to that of the linear vector field described in Sec.~\ref{sec:phasespacestruc_linear}. 
In fact if follows from general principles that all the phase structures discussed in Sec.~\ref{sec:phasespacestruc_linear} persist in the neighbourhood of the saddle (which in particular implies that one has to restrict to energies close to the energy of the saddle).  
Moreover, these phase space structures can be constructed in an algorithmic fashion  using a Poincar{\'e}-Birkhoff normal form \cite{Uzeretal02,Waalkensetal08}.
Assuming that the eigenvalues $\omega_k$, $k=2,\ldots,\dof$, are independent over the field of rational numbers (i.e. in the absence of resonances), 
the Poincar{\'e}-Birkhoff normal form yields a symplectic transformation to new (\emph{normal form}) coordinates such that 
the transformed Hamiltonian function truncated at order $n_0$ of its Taylor expansion assumes the  form 
\begin{equation} 
H_{\text{NF}}({\cal I}_1,I_2,\ldots,I_\dof)\,,
\end{equation}
where ${\cal I}_1$ and $I_k$, $k=2,\ldots,\dof$, are constants of motions which (when expressed in terms of the normal form coordinates) have the same form as in \eqref{eq:def_constants_motion}, and $H_{\text{NF}}$ is a polynomial of order $n_0/2$ in  ${\cal I}_1$ and $I_k$, $k=2,\ldots,\dof$ (note that only even orders $n_0$ of a normal form make sense).
The algorithm to compute this transformation is sketched in  Appendix~\ref{sec:NF_algorithm}. 

In terms of  the normal form coordinates the phase space structures can be defined in a manner which is virtually identical to the linear case by replacing $H_2(\vq,\vp)$ by $H_{\mathrm{NF}}({\cal I}_1,I_2,\ldots,I_\dof)$ in the definitions in Sec.~\ref{sec:phasespacestruc_linear}.  Using then the inverse of the normal form transformation allows one to construct the phase space structures in the original (`physical') coordinates. The Poincar{\'e}-Birkhoff normal form is therefore of crucial importance for the construction of the phase space structures governing reaction dynamics in a general Hamiltonian system with a saddle equilibrium point.

\section{Reduction of the N-body system}
\label{sec:Nbodyreduction}

We now want to study reaction type dynamics induced by the relative equilibria of a rotationally reduced $N$-body system.
We start by recalling the reduction procedure for the $N$-body system following
\cite{LittlejohnReinsch97}. Let $\mathbf{x}_{i}, i=1,...,N$, be the position
vectors of $N$ bodies in $\R^3$. If a set of mass-weighted Jacobi
vectors $\mathbf{s}_{i}, i=1,...,N-1$, are chosen (for an example, see Sec.~\ref{sec:triatomic}) and the center of mass
of the system is assumed to be the origin, then the three degrees of
freedom associated with overall translations are eliminated, and the kinetic energy takes a diagonal form
\begin{equation}
K=\frac{1}{2}{\sum^{N}_{i} }\mathbf{\dot{s}}_{i=1}^{2}.
\end{equation}
After a suitable choice of a body frame one can write $\mathbf{s}%
_{i}=\mR(\theta _{1},\theta _{2},\theta _{3})\mathbf{r}_{i}$ for some $%
\mR(\theta _{1},\theta _{2},\theta _{3})\in SO(3)$, where $\theta _{1},\theta
_{2},\theta _{3}$ is some set of Euler angles, and $\mathbf{r}_{i}$, $i=1,...,N-1$, are the mass-weighted Jacobi vectors in the body frame. The $\mathbf{r}_{i}$ can be expressed in $3N-6$ coordinates $q_{\mu }
$, called shape coordinates and their space is called the \emph{shape} or \emph{internal space} which we denote by $M$. Away from collinear configurations, as we will assume throughout, the translation reduced space is a fiber bundle over the shape space. 
For a conservative $N$-body system without external forces the potential energy can be viewed as  a function on the shape space (i.e. it is a function of the shape space coordinates only). The goal now is to also express  the 
kinetic energy (as far as possible) as a function of the shape space coordinates. To this end we make the following definitions.

Let $\mR\in SO(3)$ denote the rotation from the center of mass frame to the body frame and $\mathbf{L}$ denote the angular momentum
\begin{equation}
\mathbf{L}=\sum^{N}_{i=1}\mathbf{s}_{i}\times \mathbf{\dot{s}}_{i}.
\end{equation}
Then the body velocities and body angular momentum are defined, respectively, by
\begin{equation}
\mathbf{\dot{r}}_{i}=\mR^{T}\mathbf{\dot{s}}_{i},
\end{equation}
and
\begin{equation} \label{eq:def_J}
\mathbf{J}=\mR^{T}\mathbf{L} \,.
\end{equation}%

The moment of
inertia tensor $\mM (q)$ has the components
\begin{equation}
 M_{ij}(q) =  \sum^{N-1}_{k=1}(\mathbf{r}_{k}^{2}\delta _{ij}-r_{ki}r_{kj}) \,,
\end{equation}
where $\mathbf{r}_{k}=(r_{k1},r_{k2},r_{k3})$ in body coordinates, and $\mathbf{A}_{\mu }$ is the so called \emph{gauge potential} 
\begin{equation}
\mathbf{A}_{\mu }(q)=\mM^{-1}(q)\cdot \sum_{i=1}^N \left( \mathbf{r}_{i}\times \frac{\partial 
\mathbf{r}_{i}}{\partial q_{\mu }}\right)\,.
\end{equation}%
After defining the metric%
\begin{equation}
g_{\mu \nu }=\frac{\partial r_{\alpha }}{\partial q_{\mu }}\frac{\partial
r_{\alpha }}{\partial q_{\nu }}-\mathbf{A}_{\mu }\cdot \mM\cdot \mathbf{A}%
_{\nu }
\end{equation}
(where we here and in the following use the Einstein convention of summation over repeated indices which  run from 1 to $3N-6$) 
the kinetic energy becomes%
\begin{eqnarray*}
K=\frac{1}{2}(\omega +\mathbf{A}_{\mu }\dot{q}_{\mu })\cdot \mM\cdot
(\omega +\mathbf{A}_{\nu }\dot{q}_{\nu })\mathbf{+}\frac{1}{2}g_{\mu \nu }%
\dot{q}_{\mu }\dot{q}_{\nu }\,.
\end{eqnarray*}%
Here $\omega $ is the angular velocity which is the vector corresponding to
the skew-symmetric matrix $R^{T}\dot{R}$ by the natural isomorphism
\begin{equation}
\left(
\begin{array}{ccc}
0 & -\omega_3 & \omega_2 \\ 
\omega_3 & 0 & -\omega_1 \\ 
-\omega_2 & \omega_1 & 0
\end{array}
\right) \mapsto
\left(
\begin{array}{ccc}
\omega_1 \\ 
\omega_2  \\ 
\omega_3 
\end{array}
\right)\,. \label{eq:def_R}
\end{equation} 
By
using the equation
\begin{equation}
\mathbf{J}=\frac{\partial K}{\partial \omega}=\mM(\omega +\mathbf{A}_{\mu }\dot{q}_{\mu }),
\end{equation}%
the conjugate momenta of the shape space coordinates are obtained by
\begin{equation}
p_{\mu }=\frac{\partial K}{\partial \dot{q}_{\mu }}=g_{\mu \nu }\dot{q}_{\nu }+\mathbf{J}\cdot \mathbf{A}_{\mu }\,.
\end{equation}%

The fully reduced Hamiltonian is then given by
\begin{equation}  \label{Hamiltonian}
H=\frac{1}{2}\mathbf{J}\cdot \mM^{-1}\cdot \mathbf{J+}\frac{1}{2}g^{\mu \nu
}(p_{\mu }-\mathbf{J}\cdot \mathbf{A}_{\mu })(p_{\nu }-\mathbf{J}\cdot 
\mathbf{A}_{\nu })
+V(q_{1},...,q_{3N-6})\, . 
\end{equation}%
Here the first term is called \emph{rotational} or \emph{centrifugal energy} and the second one is called \emph{vibrational kinetic energy}. 

The equations of motion are obtained to be%
\begin{equation}\label{Equations_motion_old}
\dot{q}_{\mu }=\partial H/\partial p_{\mu },\ \ \ \dot{p}_{\mu }=-\partial
H/\partial q_{\mu },  \ \ \  \mathbf{\dot{J}=J\times }\frac{\partial H}{\partial \mathbf{J}} 
\end{equation}%
for $\mu =1,...,3N-6,$ \cite{KozinRobertsTennyson00}. The last equation is equivalent to \cite{KozinRobertsTennyson99}
\begin{equation}
\dot{J}_{a}=\left\{ J_{a},H\right\}, a=1,2,3, 
\end{equation}
where $\mathbf{J}=(J_{1},J_{2},J_{3}).$

The theory underlying the reduction   described above  is the orbit reduction method of Marle \cite{OrtegaRatiu04}. Let $Q$ denote the translational reduced configuration space of non-collinear configurations. Then the action of $SO(3)$ on that space is free which implies that the lifted action of $SO(3)$ on $T^{*}Q$ is also free besides being symplectic. The angular momentum can be considered as a map (a so called \emph{momentum map}) $J$ from $T^{*}Q$ to the dual space of the Lie algebra of $SO(3)$, $\mathfrak{so(3)^{*}}$, which is identified with $\R^{3}$. After writing the Hamiltonian in terms of a body fixed frame, it takes an $SO(3)$-invariant form as in (\ref{Hamiltonian}). As the (space fixed) angular momentum is constant, say $\zeta$, and the magnitute of the body angular momentum is also a constant of motion, the dynamics can be reduced to the inverse image under the angular momentum map of the body angular momentum sphere $\mathcal{O}_{\zeta}=S_{\left\Vert \zeta \right\Vert}^{2}$ which is the coadjoint orbit of the $SO(3)$ action. Then (\ref{Hamiltonian}) may be considered as a function on $J^{-1}(\mathcal{O}_{\zeta})/SO(3)$ which is symplectomorphic to the Marsden-Weinstein space $J^{-1}(\zeta)/SO(2)$, where $SO(2)$ is the isotropy subgroup of the coadjoint action of $SO(3)$ on $\mathfrak{so(3)^{*}}$.

\section{Canonical coordinates on the reduced phase space}
\label{sec:canoncoord}

\begin{figure}
\begin{center}
\raisebox{8cm}{(a)}\includegraphics[angle=0,width=7cm]{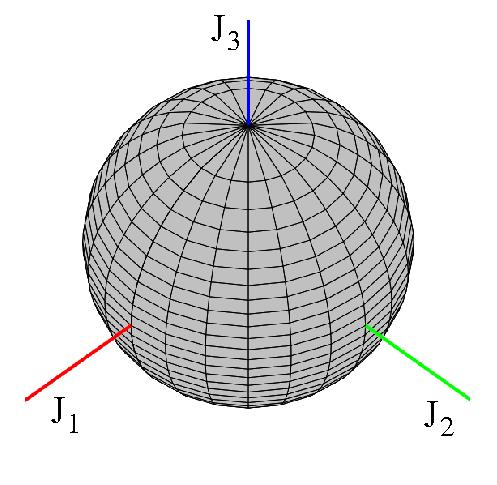}
\raisebox{8cm}{(b)}\includegraphics[angle=0,width=7cm]{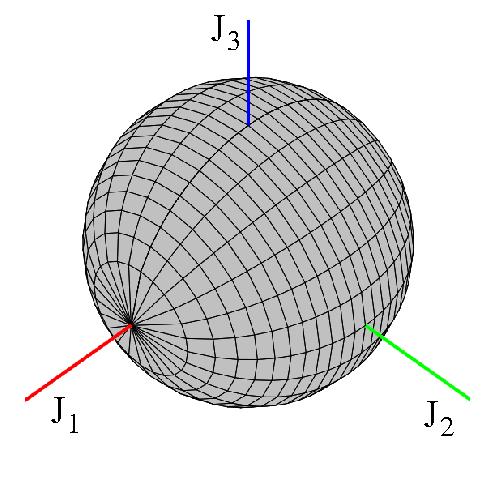}
\end{center}
\caption{\label{fig:Jsphere}
$(u,v)$ coordinate lines on the angular momentum sphere for (a) the definition of $(u,v)$ in  \eqref{Coordinates} and (b)  the definition of $(u,v)$ in  \eqref{Coordinates2}.
}
\end{figure}

It can be seen by (\ref{Hamiltonian}) that $(q_{\mu},p_{\mu},J_{a})$ are some coordinates on the reduced space where $\mu=1,..,3N-6$ and $J_{a}, a=1,2,3,$ are the components of the body angular momentum vector $\mathbf{J}$. Note here that $(q_{\mu},p_{\mu})$ are canonical variables whereas the $J_{a}$ are not canonical. If one uses coordinates $v_{\mu}=p_{\mu}-\mathbf{J}\cdot \mathbf{A}_{\mu }$ in place of $p_{\mu}$, then $(q_{\mu},v_{\mu})$ become non-canonical too \cite{LittlejohnReinsch97}  but  this way it easy to see that the reduced space is locally diffeomorphic to the product of the angular momentum sphere and the cotangent bundle of the internal space: $S_{\left\Vert \mathbf{J}\right\Vert }^{2}\times T^{\ast }M$. This is in fact a general fact for cotangent bundle reduction, known as \emph{fibration cotangent bundle reduction} \cite{OrtegaRatiu04}. It is to be noted here that
the symplectic structure on the reduced space is the sum of $\pm
1/\left\Vert \mathbf{J}\right\Vert $ times the volume form of the sphere, the canonical symplectic structure on $T^{\ast }M$ and some magnetic terms \cite{OrtegaRatiu04}. The
magnetic terms come from the gauge potential, and are directly related to the non-Euclidean
structure of the shape space. The goal now is to use some canonical coordinates $%
u,v$ on $S_{\left\Vert \mathbf{J}\right\Vert }^{2}$ instead of the coordinates $J_{a}$. The related symplectic structures on the coadjoint orbit $S_{r}$, where $\
r=\left\Vert \mathbf{J}\right\Vert ,$ are%
\begin{equation} \label{eq:def_w_pm}
w_{\pm }=\pm \frac{1}{r}dA
\end{equation}%
where $dA$ is the volume form on the sphere, or in any coordinates $u,v$ on $%
S^2_{r}$%
\begin{equation}
w_{\pm }=\pm \frac{1}{r}\sqrt{\det \left[ g_{uv}\right] }du\wedge dv\,,
\end{equation}%
where $\left[ g_{uv}\right] $ is the matrix of the induced Riemannian metric on $S_{r}^{2}$ by
the Euclidean metric on $\R^{3}$. The canonical coordinates with respect to this symplectic structure are those that give%
\[
w_{\pm }=\pm du\wedge dv\,.
\]
One possible choice to achieve this is \cite{Deprit67}%
\begin{equation} \label{Coordinates}  
J_{1} =\sqrt{r^{2}-v^{2}}\cos u,\,
J_{2} =\sqrt{r^{2}-v^{2}}\sin u,\, J_{3} =v \,,
\end{equation}%
where $0\leq u\leq2\pi, -r\leq v\leq r$  (see Fig.~\ref{fig:Jsphere}(a)). Let $u,v$ be some canonical coordinates for the positive signed symplectic structure $w_{+}$ and $q_{\mu},p_{\mu}, 1\leq \mu \leq 3N-6$, be any choice of shape coordinates and their conjugate momenta, respectively. In terms of these coordinates, after relabeling $u=q_{0}, v=p_{0}$, and setting $\vz=(z_{i})=(q_{i},p_{i}), i=0,...,3N-6$, the
equations of motion read%
\begin{equation}\label{Equations_motion_1}
\dot{z}_{i}=\left\{ z_{i},H\right\} 
\end{equation}%
or equivalently %
\begin{equation} \label{Equations_motion}
\dot{q}_{i}=\partial H/\partial p_{i},\ \ \ \dot{p}_{i}=-\partial H/\partial
q_{i}  
\end{equation}%
for $i=0,...,3N-6$, where  the Poisson bracket on the reduced space has the standard form%
\begin{equation}
\left\{ f,g\right\} =  \sum_{i=1}{3N-6}\frac{\partial f}{\partial q_{i}}\frac{\partial g}{\partial p_{i}}-\frac{\partial f}{\partial p_{i}}\frac{\partial g}{\partial q_{i}}.
\end{equation}%

This canonical form of the equations of motion has many advantages. 
Although the
coordinates $p_{\mu }$, $\mu =1,...,3N-6$, are gauge dependent, for
examining the local behaviour of the reduced  system they can be used to study, for example, the stability
of relative equilibria \cite{KozinRobertsTennyson99, KozinRobertsTennyson00}. We will  use the canonical equations of motion to construct the phase space structures governing reaction type dynamics 
associated with saddle type relative equilbria in the reduced system from a Poincar{\'e}-Birkhoff normal form
in the same way as described in Sec.~\ref{sec:phasespacestruc}.

Relative equilibria are trivial solutions of the equations of motion (\ref{Equations_motion_old}) or equivalently (\ref{Equations_motion}). Using (\ref{Equations_motion_old}) one finds that the relative equilibria are the solutions of the reduced equations \cite{KozinRobertsTennyson00}
\begin{eqnarray}
\mathbf{J}\times \left( \mM^{-1}\cdot \mathbf{J}\right)  &=&0, \\
p_{\mu } &=&\mathbf{J}\cdot \mathbf{A}_{\mu }, \\
\frac{\partial }{\partial q_{\mu }}(\mathbf{J}\cdot \mM^{-1}\cdot \mathbf{J+}%
V(q)) &=&0.
\end{eqnarray}
This implies that at relative equilibria the body angular momentum $\vJ$ is parallel to a principal axis, i.e. 
the molecule is rotating about one of its principal axis.  Using this fact the relative equilibria can be found from the critical points of the \emph{effective potential}
\begin{equation}\label{eq:def_V_eff}
V_{\text{eff}}(q)  \mathbf{J}\cdot \mM^{-1}\cdot \mathbf{J+} V(q)\,,
\end{equation}
where  $\mathbf{J}$ is a fixed vector of a given modulus $r$ parallel to a chosen principal axis. Some of these relative equilibria are related to the  equilibria of the system with zero angular momentum. In fact for a triatomic molecule without any discrete symmetries, six families of relative equilibria are born out of any generic equilibrium if the modulus of the angular momentum is increased from zero.
Here `generic' means that the rotations act on the equilibrium configuration freely (this excludes collinear configurations) \cite{KozinRobertsTennyson00}. We will follow this procedure to find relative equilibria in the example given in Sec.~\ref{sec:triatomic}.

Given a relative equilibrium we can determine its (linear) stability using the equations of motion
 \eqref{Equations_motion}. If the stability is of saddle type (as defined in the introduction) we can 
 apply the Poincar{\'e}-Birkhoff normal form procedure described in Appendix~\ref{sec:NF_algorithm} to the equations of motion \eqref{Equations_motion}
 in order to construct the phase structures governing reaction type dynamics as described in Sec.~\ref{sec:phasespacestruc}. 
To this end we note that  $(u,v)$ defined as in \eqref{Coordinates} are singular at the north pole and south pole $\vJ=(0,0,\pm r)$ of the angular momentum sphere. 
The coordinates are therefore not useful to study relative equilibria associated with rotations about the third principal axis  \cite{CushmanBates}.  
In order to study these relative equilibria one can redefine $(u,v)$ according to
\begin{equation} \label{Coordinates2}
J_{1} =v,\,
J_{2} =\sqrt{r^{2}-v^{2}}\sin u,\, J_{3} =\sqrt{r^{2}-v^{2}}\cos u
\end{equation}%
in which case the coordinate singularities  are at the points $\vJ=(0,\pm r,0)$ on the angular momentum sphere (see Fig.~\ref{fig:Jsphere}(b)).


\section{Examples}
\label{sec:examples}

In the following we illustrate the approach above for the limiting case of a rigid molecule and the 
case of a triatomic molecule. 

\subsection{The limiting case of a rigid molecule}

\begin{figure}
\begin{center}
\raisebox{7cm}{(a)}\includegraphics[angle=0,width=7cm]{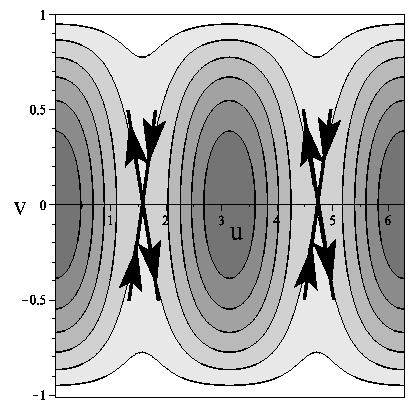}
\raisebox{7cm}{(b)}\includegraphics[angle=0,width=7cm]{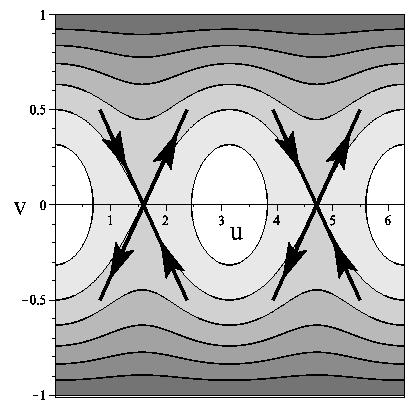}
\end{center}
\caption{\label{fig:rigdbody_equipotentials_uv}
Contours of the reduced rigid body Hamiltonian in the $(u,v)$ coordinate plane with $(u,v)$ defined according to \eqref{Coordinates} (a) and with $(u,v)$ defined  \eqref{Coordinates2} (b). 
The energy increases from light to dark shading. The bold lines and arrows indicate the eigenvectors and the corresponding  directions of the Hamiltonian  flow for the saddles.
The parameters are $M_1=1$, $M_2=2$, $M_3=3$ and $r=1$. 
}
\end{figure}

In the limiting case of a rigid molecule the internal degrees of freedom are frozen, i.e. $\dot{q}_{\mu }=\dot{q}_{\mu }=0$, $1\leq \mu
\leq 3N-6$. In this case, one can neglect the potential in  \eqref{Hamiltonian} and the Hamiltonian reduces to 
\begin{equation} \label{eq:Hrigidbody_gen}
H=\frac{1}{2}\mathbf{J}\cdot \mM^{-1}\cdot \mathbf{J}
\end{equation}%
which describes the free rotation of a rigid body. Choosing the body frame to coincide with the principal axes, 
$\mM$ (and its inverse $\mM^{-1}$) becomes diagonal and \eqref{eq:Hrigidbody_gen} becomes%
\begin{equation}
H=\frac{1}{2}\Big(  \frac{J_{1}^{2}}{M_{1}}  +    \frac{J_{2}^{2}}{M_2} +  \frac{J_{3}^{2}}{M_3}   \Big)
\end{equation}%
where $M_{1}$, $M_{2}$ and $M_{3}$ are the principal moments of inertia. In this case the angular momentum part of  \eqref{Equations_motion_old}  become the classical Euler equations 
\cite{Marsden92}. 
In terms of the coordinates $(u,v)$ defined in \eqref{Coordinates} these equations become
\begin{equation} \label{Euler}
\dot{u}=\partial H/\partial v, \ \ \ \dot{v}=-\partial H/\partial u \,.
\end{equation}%
Here the role of $v$ is seen to be a 
momentum conjugate to $v$
but when we choose the other symplectic form $\omega ^{-}=-\frac{1}{r}%
dA$ (see \eqref{eq:def_w_pm}), then the roles of $u$ and $v$ are reversed. If we choose the coordinates %
(\ref{Coordinates}), then the Hamiltonian becomes%
\begin{eqnarray*}
H(u,v) =&&\frac{1}{2}   \Big(   \frac{(r^{2}-v^{2})\cos ^{2}u}{M_1} +
  \frac{(r^{2}-v^{2})\sin ^{2}u}{M_2}+   \frac{v^{2}}{M_3}  \Big).
\end{eqnarray*}
So, by (\ref{Euler}) the equations of motion are obtained to be%
\begin{eqnarray*}
\dot{u}&=&- \frac{v\cos ^{2}u}{M_1}  -    \frac{v\sin ^{2}u}{M_2}  +  \frac{v}{M_3}\,, \\
\dot{v}&=&   \frac{(r^{2}-v^{2})\sin u\cos  u}{M_1}  -  \frac{(r^{2}-v^{2})\sin u\cos u}{M_2}\,.
\end{eqnarray*}%
In the case of the Euler top where the  moments of inertia are mutually different, i.e. where we can assume without restriction that
$M_{1}<M_{2}<M_{3}$, there are six relative equilibria given by the points $(\pm r,0,0)$, $(0,\pm r,0)$ and $(0,0,\pm r)$  on the $\vJ$ sphere. 
The dynamics near the first two can be studied in terms of the coordinates $(u,v)$ defined according to \eqref{Coordinates} where they correspond to 
the points $(0,0)$ (for $\vJ=(r,0,0)$), $(\pi,0 )$ (for $\vJ=(-r,0,0)$),  $(\pi /2,0)$ (for $\vJ=(0,r,0)$) and $(3\pi/2,0)$ (for $\vJ=(0,-r,0)$) (see Fig.~\ref{fig:rigdbody_equipotentials_uv}(a)).
The dynamics near the relative equilibria $\vJ=(0,0,\pm r)$ (and again $\vJ=(0,\pm r,0)$) can be studied in terms of $(u,v)$  defined according to \eqref{Coordinates} where they correspond to the points 
the points $(0,0)$ (for $\vJ=(0,0,r)$), $(\pi,0)$ (for $\vJ=(0,0,-r)$),  $(\pi/2,0)$ (for $\vJ=(0,r,0)$) and $(3\pi/2,0)$ (for $\vJ=(0,-r,0)$) (see Fig.~\ref{fig:rigdbody_equipotentials_uv}(b)).
The reduced Hamiltonian has local maxima at $\vJ=(\pm r,0,0)$ which correspond to rotations in either direction about the principal axis with the smallest moment of inertia, and minima at $\vJ=(0,0,\pm r)$ which correspond to rotations in either direction about the principal axis with the largest moment of inertia. The rotations about the principal axis with middle moment of inertia corresponding to $\vJ=(0,\pm r,0)$ are unstable. The corresponding relative equilibria are of saddle type.  The eigenvalues associated with the linearized vector field are 
$\pm \sqrt{M_1 M3 (M_2-M_1)(M_3-M_2)}/(M_1 M_2 M_3)$. The eigenvectors are shown in Fig.~\ref{fig:rigdbody_equipotentials_uv}. Since the reduced rigid body motion has one degree of freedom the reaction dynamics associated with these saddles is trivial (see the remarks at the end of Sec.~\ref{sec:phasespacestruc_linear}). One can divide the phase space of the reduced system into the regions $v<0$  and $v>0$ in terms of the coordinates \eqref{Coordinates}  or equivalently in the regions $\vert u \vert <\pi/2  $ and $\vert u \vert >\pi/2$ in terms of the coordinates \eqref{Coordinates2} (where we used the periodicity of $u$ in the last case; see Fig.~\ref{fig:rigdbody_equipotentials_uv}). For energies below $r^2/(2M_2)$ which is the energy of the saddle one cannot reach one region from the other. For energies above $r^2/(2M_2)$ however this  becomes possible. 

\subsection{Triatomic molecules: HCN}
\label{sec:triatomic}

\begin{figure}
\begin{center}
\includegraphics[angle=0,width=7cm]{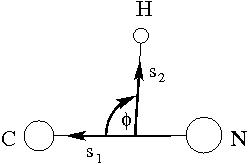}
\end{center}
\caption{\label{fig:HCN_conf}
Definition of the Jacobi vectors $\mathbf{s}_{1}$ and $\mathbf{s}_{2}$ and the corresponding angle $\phi$ (see \eqref{eq:def_s1s2} and \eqref{eq:def_rhophi}) for the HCN molecule.  Carbon is associated with the first mass, nitrogen is associated with the second mass and hydrogen is associated with the third mass.
}
\end{figure}

For a second example, we consider a triatomic molecule. We will use the $xxy$ gauge and Dragt's coordinates as follows \cite{LittlejohnReinsch97}. Let $\mathbf{x}_{1}, \mathbf{x}_{3},\mathbf{x}_{3}$ be the position
vectors of $3$ bodies in $\R^3$. If  mass-weighted Jacobi
vectors $\mathbf{s}_{1},\mathbf{s}_{1}$ are chosen as
\begin{equation} {\label{eq:def_s1s2}}
\mathbf{s}_{1}=\sqrt{\mu_1}(\mathbf{x}_{1}-\mathbf{x}_{3})\,,\quad
\mathbf{s}_{2}=\sqrt{\mu_2}(\mathbf{x}_{2}-\frac{m_1\mathbf{x}_{1}+m_3\mathbf{x}_{3}}{m_1+m_3}),
\end{equation}
where 
\begin{equation}\label{eq:def_reduced_masses}
\mu_1=\frac{m_1m_3}{m_1+m_3} \text{ and } \mu_2=\frac{m_2(m_1+m_3)}{m_1+m_2+m_3}
\end{equation}
are the reduced masses then the corresponding Jacobi coordinates $(\rho_1,\rho_2,\phi)$ are defined by
\begin{equation*}      \label{eq:def_rhophi}        
\rho_1=\|\mathbf{s}_{1}\|\,,\quad \rho_2=\|\mathbf{s}_{2}\|\,, \quad \mathbf{s}_{1}\cdot\mathbf{s}_{2}=\rho_1\rho_2\cos\phi\,,
\end{equation*}
where $0\leq\phi\leq\pi$. 
From the Jacobi coordinates one can define the coordinates
\begin{equation}
(w_1,w_2,w_3) = (\rho_1^{2}-\rho_2^{2}, 2\rho_1\rho_2\cos\phi, 2\rho_1\rho_2\sin\phi )\,,
\end{equation}
where $w_1,w_2\in \R$ and $w_3\ge 0$.
 Dragt's coordinates $(\omega,\chi,\psi )$ are now polar coordinates in the $(w_1,w_2,w_3)$ coordinate space:
\begin{equation}
(w_1,w_2,w_3) = (\omega\cos\chi\cos\psi, \omega\cos\chi\sin\psi ,\omega\sin\chi)\,,
\end{equation}
where $\omega>0, 0<\chi\le\pi/2, 0\le\psi\le2\pi.$ 
Note that $\xi$ is the latitude, not the colatitude. 
The inertia and metric tensors are diagonal in Dragt's coordinates  \cite{LittlejohnReinsch97}:
\begin{equation}
\mM =\left[ 
\begin{array}{ccc}
\omega \sin ^{2}\frac{\chi }{2} & 0 & 0 \\ 
0 & \omega \cos ^{2}\frac{\chi }{2} & 0 \\ 
0 & 0 & \omega %
\end{array}%
\right]
\end{equation}
and
\begin{equation}
\left[ g^{\mu\nu}\right] =\left[ 
\begin{array}{ccc}
4\omega & 0 & 0 \\ 
0 & \frac{4}{\omega} & 0 \\ 
0 & 0 & \frac{4}{\omega \cos^2\chi}%
\end{array}%
\right]\,.
\end{equation}
The gauge potential becomes
\begin{equation}
A_{\omega }=A_{\chi }=(0,0,0),\ \ \ A_{\psi }=(0,0,-\frac{1}{2}\sin \chi ).
\end{equation}%
Hence, if the coordinates (\ref{Coordinates}) are used then
after relabeling $u=q_{0},v=p_{0},\omega =q_{1},\chi =q_{2},\psi =q_{3}$, the reduced  Hamiltonian becomes 
\begin{eqnarray*}
H(q,p)&=&\frac{(r^{2}-p_{0}^{2})\cos ^{2}q_{0}}{2q_{1}\sin ^{2}\frac{q_{2}}{2}}+\frac{(r^{2}-p_{0}^{2})\sin ^{2}q_{0}}{2q_{1}\cos ^{2}\frac{q_{2}}{2}}+\frac{p_{0}^{2}}{2q_{1}}+2q_{1}p_{1}^{2}+\frac{2p_{2}^{2}}{q_{1}}+\frac{(2p_{3}+p_{0}\sin{q_{2}})^2}{2q_{1}\cos^2 q_{2}}\\
&&+V(q_{1},q_{2},q_{3})\,,
\end{eqnarray*}
or if the coordinates (\ref{Coordinates2}) are used one has
\begin{eqnarray*}
H(q,p)&=&\frac{p_{0}^{2}}{2q_{1}\sin^{2}\frac{q_{2}}{2}}+\frac{(r^{2}-p_{0}^{2})\sin^{2}q_{0}}{2q_{1}\cos^{2}\frac{q_{2}}{2}}+\frac{(r^{2}-p_{0}^{2})\cos^{2}q_{0}}{2q_{1}}+2q_{1}p_{1}^{2}+\frac{2p_{2}^{2}}{q_{1}}\\
&&+\frac{(2p_{3}+\sqrt{r^{2}-p_{0}^{2}}\cos q_{0}\sin{q_{2}})^2}{2q_{1}\cos^{2}q_{2}}+V(q_{1},q_{2},q_{3}).
\end{eqnarray*}
\rem{
Relative equilibria are the critical points of the reduced Hamiltonian but as the coordinates are local, one has to slightly modify the coordinates in order to find all critical points. Also, at a relative equilibrium $\mathbf{J}$ must be a principal axis. So in Dragt's coordinates $\mathbf{J}$ has values $(\pm r,0,0), (0,\pm r,0), (0,0,\pm r)$. Then the relative equilibria are the critical points of         
\begin{equation}
\Veff=\mathbf{J}\cdot \mM^{-1}\cdot \mathbf{J}+V(q_{1},q_{2},q_{3}).
\end{equation}
In general the potential function is a function of the distances between particles, i.e., $V=V(d_{ij})$ where $d_{ij}$ is the distance between the $i$-th  and $j$-th particles. In Dragt's coordinates these distances are
\begin{eqnarray*}
d_{13} &=&\frac{1}{\sqrt{2\mu_1}}\sqrt{q_{1}+q_{1}\cos q_{2}\cos q_{3}}, \\
d_{12} &=&\sqrt{\frac{\mu_1}{2m_1^2}(q_1+q_1\cos q_{2}\cos q_{3})+\frac{1}{2\mu_2}(q_1-q_1\cos q_{2}\cos q_{3})-\frac{\sqrt{\mu_1}}{m_1\sqrt{\mu_2}}q_{1}\cos
q_{2}\sin q_{3}}, \\
d_{23} &=&\sqrt{\frac{\mu_1}{2m_3^2}(q_1+q_1\cos q_{2}\cos q_{3})+\frac{1}{2\mu_2}(q_1-q_1\cos q_{2}\cos q_{3})+\frac{\sqrt{\mu_1}}{m_3\sqrt{\mu_2}}q_{1}\cos
q_{2}\sin q_{3}}.
\end{eqnarray*}
So, after finding the critical points of $\Veff$, the Hessian of $H$
is 
tested at critical points. 
} 

\begin{figure}
\begin{center}
\raisebox{5cm}{(a)}\includegraphics[angle=0,width=7cm]{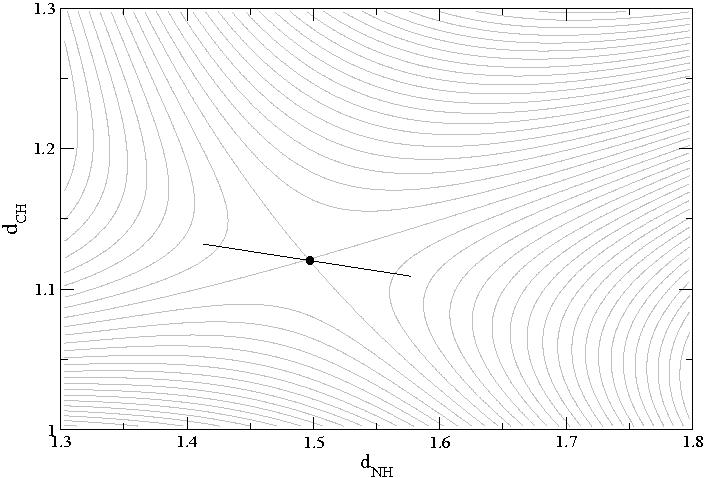}
\raisebox{5cm}{(b)}\includegraphics[angle=0,width=7cm]{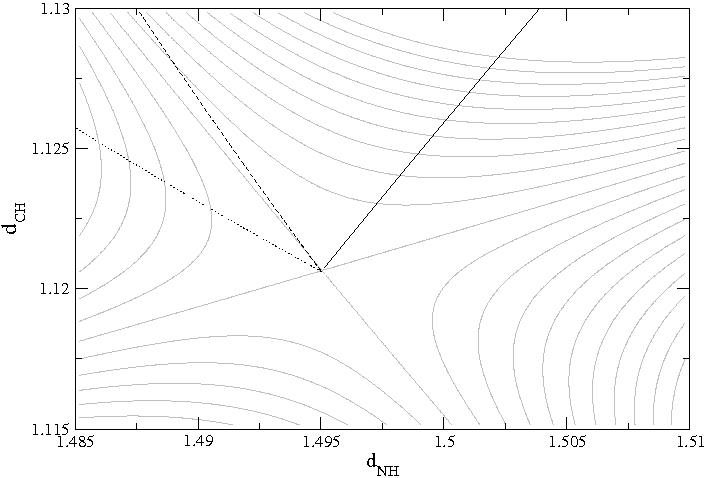}
\end{center}
\caption{\label{fig:HCN_potentials}
Potential energy lines (grey)  of the Murrell, Carter and Halonen \cite{MurrellCarterHalonen82} HCN potential in the section 
$d_{\text{CN}} = 1.1394
$ \AA\, together with (a) the projection of the real eigenvectors of the saddle equilibrium of the system with vanishing angular momentum and (b) the path traced by the relative equilibrium under variation of $r$ projected to this plane  for rotations about the first  (full line), second (dashed line) and third  (dotted line) principal axis. 
}
\end{figure}

As a concrete example we consider the HCN molecule where we use the potential by Murrell, Carter and Halonen \cite{MurrellCarterHalonen82}.
The molecule exists in the form of two isomers which correspond to the collinear configurations HCN (hydrogen cyanide; the corresponding equilibrium has energy -13.5914 eV) and CNH (hydrogen isocyanide; the corresponding equilibrium has energy -13.1065 eV). For zero angular momentum, the reaction  dynamics of the isomerization from HCN to CNH or vice versa is induced by a saddle equilibrium point which corresponds to a triangular (noncollinear) configuration (and has an energy of -12.0827 eV).  The internuclear distances corresponding to this saddle are 
\begin{equation}
d_{\text{CN}} = 1.1394
\text{ \AA} \,,\quad
d_{\text{CH}} = 1.1206
\text{ \AA} \,,  \quad
d_{\text{NH}} = 1.4950
\text{ \AA}\,.
\end{equation}
The eigenvalues of the matrix associated with the linearized vector field at this equilibrium are 
$\pm 589.31
\,\ui$, $\pm 418.69
\, \ui$ and $\pm 213.08
$  (in units of inverse pico seconds) indicating that it is of saddle$\times$center$\times$center stability type.

We assign the Dragt's coordinates by identifying the first body with carbon, the second body with nitrogen and the third body with hydrogen (see Fig.~\ref{fig:HCN_conf}). The internuclear distances are then given by
\begin{eqnarray*}
d_{CN} &=&\frac{1}{\sqrt{2\mu_1}}\sqrt{q_{1}+q_{1}\cos q_{2}\cos q_{3}}, \\
d_{CH} &=&\sqrt{\frac{\mu_1}{2m_1^2}(q_1+q_1\cos q_{2}\cos q_{3})+\frac{1}{2\mu_2}(q_1-q_1\cos q_{2}\cos q_{3})-\frac{\sqrt{\mu_1}}{m_1\sqrt{\mu_2}}q_{1}\cos
q_{2}\sin q_{3}}, \\
d_{NH} &=&\sqrt{\frac{\mu_1}{2m_3^2}(q_1+q_1\cos q_{2}\cos q_{3})+\frac{1}{2\mu_2}(q_1-q_1\cos q_{2}\cos q_{3})+\frac{\sqrt{\mu_1}}{m_3\sqrt{\mu_2}}q_{1}\cos
q_{2}\sin q_{3}}\,,
\end{eqnarray*}
where $m_1$, $m_2$ and $m_3$ are the masses of carbon, nitrogen and hydrogen, respectively, and $\mu_1$ and $\mu_2$ are the reduced masses defined according to \eqref{eq:def_reduced_masses}.
The internuclear distances can also be viewed a coordinates on the shape space. The equipotential lines of the Murrell-Carter-Halonen potential intersected with the plane $d_{\text{CN}} = 1.1394$ (which contains the saddle) are shown in Fig.~\ref{fig:HCN_potentials}(a). 
In Fig.~\ref{fig:HCN_potentials}(a) also the projection of the eigenvectors corresponding to the real eigenvalues of the saddle are shown. Due to the time reversal symmetry of the system without angular momentum they project to the same line.

\begin{figure}
\begin{center}
\includegraphics[angle=0,width=10cm]{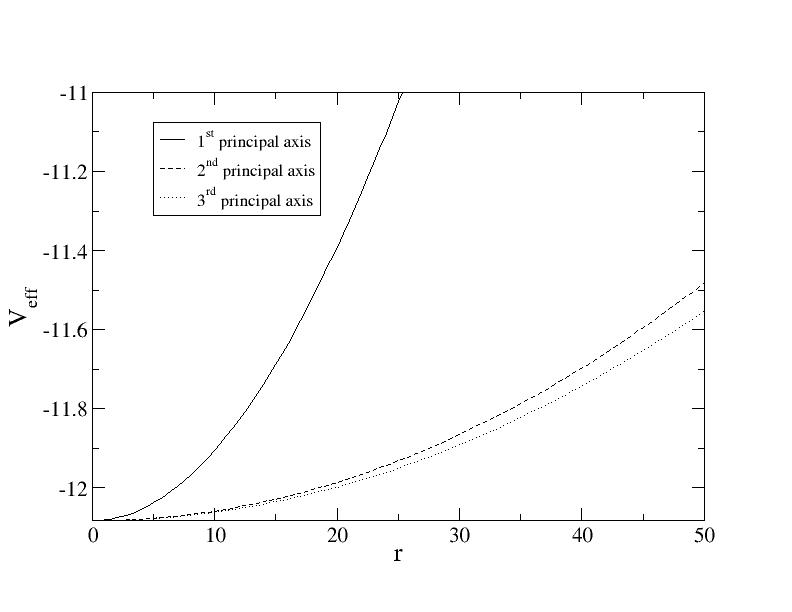}
\end{center}
\caption{\label{fig:RE_energies}
Energies  of the relative equilibria as a function of the modulus of the angular momentum $r$. The energy is measured in eV and $r$ is measured in units of $\hbar$. For fixed $r$, the relative equilibrium involving rotations about the principal axis with smallest (highest) moment of inertia has highest (smallest) energy.
}
\end{figure}

\begin{figure}
\begin{center}
\includegraphics[angle=0,width=10cm]{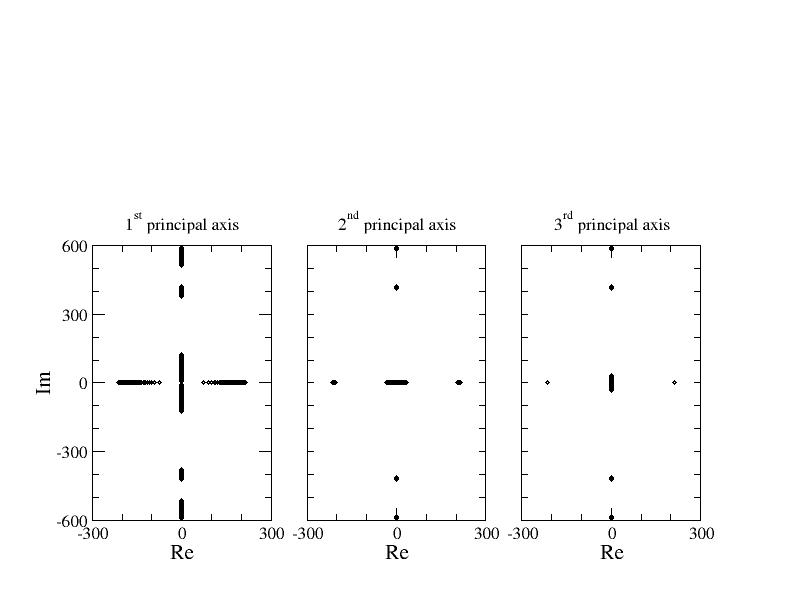}
\end{center}
\caption{\label{fig:eigenvalues}
Paths traced by the eigenvalues (in units of inverse pico seconds) of the linearized vector fields as the modulus of the angular momentum $r$ is varied  for rotations about the first principal axis (left panel), the second principal axis (middle panel) and the third principal axis (right panel). The ranges for $r$ are about the same as in Fig.~\ref{fig:RE_energies}. 
}
\end{figure}

For the zero angular momentum saddle, 
the moment of inertia tensor has the diagonal elements 
$[1.1645
, 8.6470
,9.8115
]$ (in the units of \AA$^2\times$eV). Due to the similar masses of carbon and nitrogen the values of the second and third moment of inertia are quite similar. 
We consider the six relative equilibria resulting from this saddle as the modulus $r$ of the angular momentum is increased from zero (see Sec.~\ref{sec:canoncoord}).
Since the directions of rotations about either principal axis are related by symmetry there are effectively three families. We numerically compute  these relative equilibria as a function of $r$  from the critical points of the effective potential $V_{\text{eff}}(q_1,q_2,q_3)$ in \eqref{eq:def_V_eff} where the angular momentum $\mathbf{J}$ is $(r,0,0)$, $(0,r,0)$ or $(0,0,r)$, respectively,
using  a Newton procedure. We show the energies of these relative equilibria as a function of $r$ in Fig.~\ref{fig:RE_energies}.
Since for a rigid body, the rotation about the principal axis with middle moment of inertia is unstable 
it is to be expected, that at least for small values of $r$, the two of the relative equilibria involving rotations about the first and the third principal axis are  saddle$\times$center$\times$center$\times$centers and the relative equilibrium involving rotations about the second principal axis is a saddle$\times$saddle$\times$center$\times$center (which has two pairs of real eigenvalues and two complex conjugate pairs of imaginary eigenvalues).
Linearizing the vector fields at the relative equilibria we find that this is indeed the case. In Fig.~\ref{fig:eigenvalues} we show the disposition of the eigenvalues in the complex plane as a function of $r$ for the three cases.
For $r\to0$,
 the eigenvalues reduce in each case to the eigenvalues of the equilibrium in the system without angular momentum plus a double eigenvalue at zero.
 
 Figure~\ref{fig:HCN_potentials} shows the projections of the paths traced by the relative equilibria under variation of $r$ to the $(d_{\text{CH}},d_{\text{NH}})$ plane. One finds that in this projection
 the relative equilibrium corresponding to rotations about the first principal axis is moving much faster than the other two relative equilibria as $r$ is varied.  Note that the position of the relative equilibria on the angular momentum sphere does not change under variation of $r$.
 
\begin{figure}
\begin{center}
\raisebox{5cm}{(a)}\includegraphics[angle=0,width=7cm]{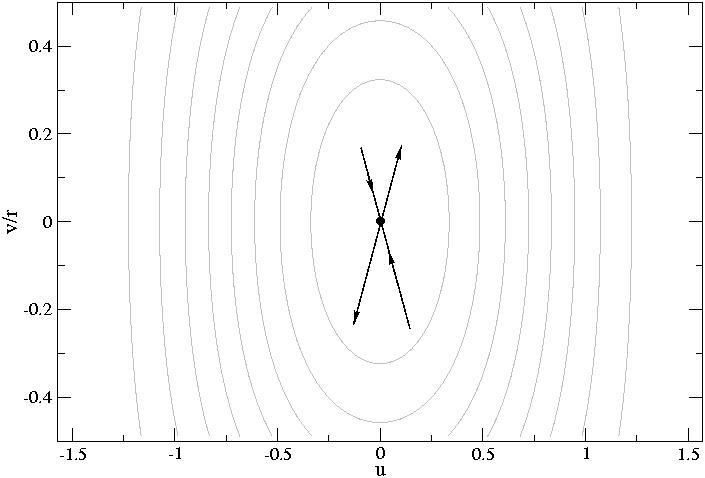}
\raisebox{5cm}{(b)}\includegraphics[angle=0,width=7cm]{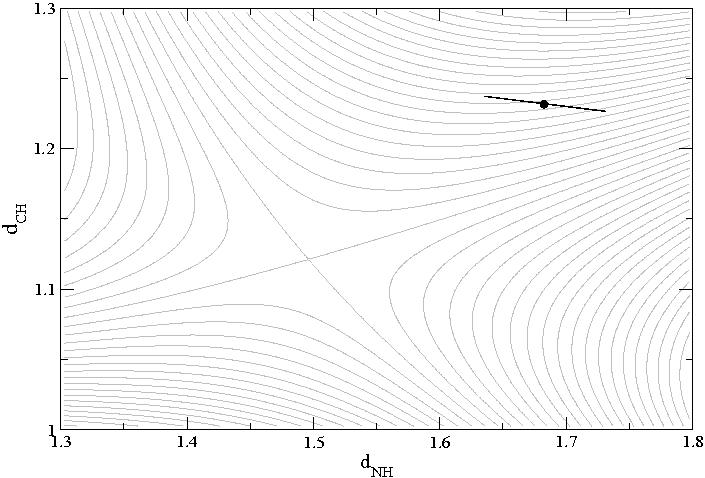}\\
\raisebox{5cm}{(c)}\includegraphics[angle=0,width=7cm]{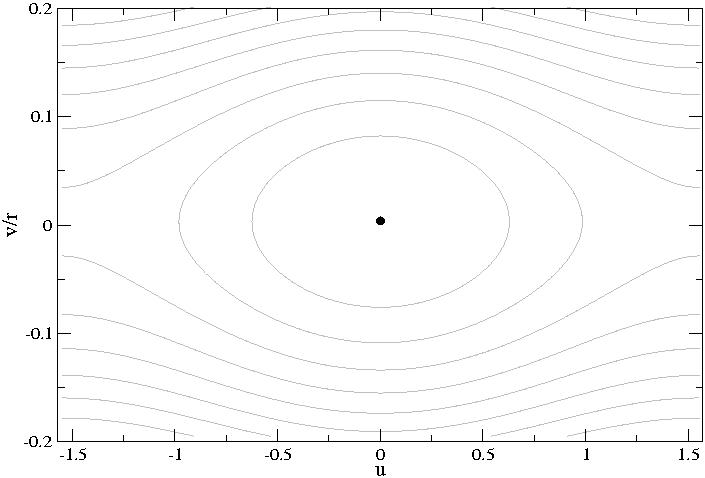}
\raisebox{5cm}{(d)}\includegraphics[angle=0,width=7cm]{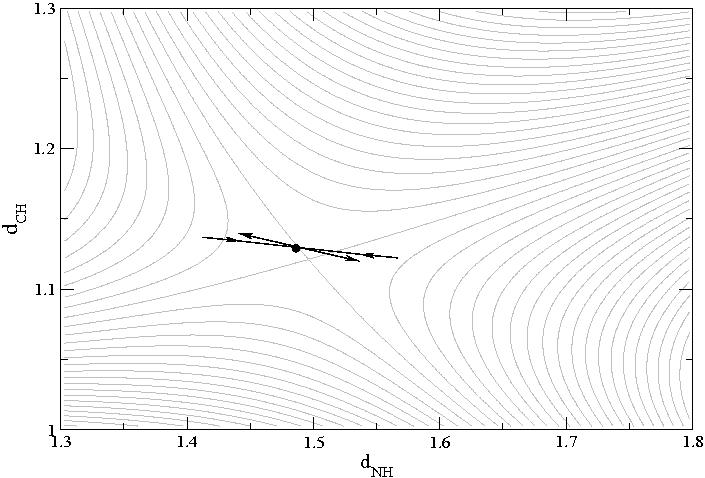}
\end{center}
\caption{\label{fig:eigenvectors}
Saddle type relative equilibria and the corresponding real eigenvectors projected to the $(u,v)$ plane and $(d_{\text{CH}},d_{\text{NH}})$ plane, respectively for rotations about the first principal axis with $r=33.5\,,\hbar$ ((a) and (b)), and for rotations about the third principal axis  with $r=50\,,\hbar$ ((c) and (d)).
}
\end{figure}

 The saddle type relative equilibria corresponding to rotations about the first and the third principal axis induce reaction type dynamics. The dynamics and its physical implications will be discussed in another publications. We here restrict ourselves to a brief discussion of the reaction paths in the limit where the energy approaches the energy of the relative equilibrium from above (see the discussion in Sec.~\ref{sec:phasespacestruc}). In this limit the reaction paths reduce to the one-dimensional stable and unstable manifolds of the relative equilibria. At the relative equilibria these lines are tangent to the eigenvectors corresponding to the real eigenvalues of the linearization of the reduced system. We therefore show pictures of the pairs of real eigenvectors of the relative equilibria in Fig.~\ref{fig:eigenvectors}. For the rotation about the first principal axis, the projections of the eigenvectors corresponding to positive and negative eigenvalues to the $(d_{\text{CH}},d_{\text{NH}})$ plane are almost identical (similar to the angular momentum zero case). Their projections to the $(u,v)$ plane however, are quite different (see Figs.~\ref{fig:eigenvectors}(a) and (b)). 
As the example indicates the relative equilibrium (in its projection to the shape space) might be quite far away from the  position of the corresponding equilibrium of the system without angular momentum, and hence the reaction paths change quite considerably as a function of $r$. 
In the case of rotations about the third principal axis the projections of the eigenvectors of the two real eigenvalues to the $(d_{\text{CH}},d_{\text{NH}})$ plane are quite different whereas their projections to the $(u,v)$ plane is zero. The reason for the latter is related to the fact that in the case of rotation about the third principal axis the motion is planar (in the $xxy$ gauge we are using the angular momentum vector is perpendicular to the plane spanned by the three bodies for rotations about the third principal axis). Although this relative equilibrium (in its projection to the shape space) is still relatively close to the corresponding equilibrium of the system without angular momentum the different directions of the eigenvectors indicate that the forward and backward reaction paths might globally be quite different for the system with and without angular momentum.

\section{Conclusions and Outlook }
\label{sec:conclusions}

In this paper we gave an explicit formalism to locally construct canonical coordinates in rotational symmetry reduced $N$-body systems. 
This allows one to study relative equilibria in such systems and 
use Poincar{\'e}-Birkhoff normal form to determine the phase space structures that govern reaction type dynamics near relative equilibria of saddle stability type.  
We briefly  illustrated the formalism by applying it to the saddle relative equilibria which induce the isomerization reaction dynamics of HCN/CNH with rotation. 
We here restricted ourselves to the study of the reaction paths near the saddle relative equilibria. A more detailed study of the isomerization dynamics which includes a Poincar{\'e}-Birkhoff normal computation will be presented elsewhere.
As an outlook we give a short list of open questions and future directions related to the work presented in this paper. 
A question that we did not discuss is the reconstruction problem, i.e. the study of what the corresponding motions of the reduced system are in the  full original system. Especially from the perspective of reaction dynamics this is an interesting and important problem. 
The formalism introduced in this paper is not restricted to the study of relative equilibria of saddle type. Recently also the  relevance of higher rank saddles for reaction type dynamics has  been studied \cite{EzraWiggins09,Halleretal11}. In the HCN/CNH isomerization problem with angular momentum a saddle$\times$saddle$\times$center$\times$center relative equilibrium (and its symmetric partner) is presented by the rotation about the principal axis with the middle moment of inertia  (and the two directions of rotation). Interestingly, for a fixed value of the modulus of the angular momentum, these rank 2 saddle relative equilibria are quite close in energy to the saddle relative equilibria corresponding to motion about the principal axis with the largest moment of inertia    (see Fig.~\ref{fig:RE_energies}).  It would be interesting to study the interplay between the phase space structures associated with the rank two saddles and usual saddles and their influence on the global dynamics. 
In \cite{WaalkensBurbanksWiggins05,WaalkensBurbanksWiggins05c,WaalkensBurbanksWiggins05b} (see also \cite{EzraWaalkensWiggins2009}) a procedure based on transition state theory  to determine the microcanonical volume of initial conditions that lead to reactive trajectories has been introduced. The formalism presented in this paper allows one to study such reactive volumes for rotational symmetry reduced systems. An interesting question for applications would be to study how this  reactive volume changes when  the modulus of the angular momentum is varied.
The formalism presented in this paper at first only applies to relative equilibria of noncollinear configurations. We remark however that it can be quite simply adapted  to also include  collinear configurations. For this purpose, one can consider the Hamiltonian and coordinates in 
\cite{KozinRobertsTennyson00} and choose suitable coordinates in place of the angular momentum.
In this paper we did not discuss the quantum reaction dynamics of rotating molecules. In \cite{SchubertWaalkensWiggins06,Waalkensetal08}  the Weyl symbol calculus has been used to develop a quantum mechanical analogue of the Poincar{\'e}-Birkhoff normal form which provides an efficient procedure to compute quantum reaction rates and the related Gamov-Siegert resonances. The same procedure can in principle also be applied to the relative equilibria of rotational symmetry reduced molecules. However, the reduction of the quantum $N$-body system has some differences to the classical case. For example, for the quantum three body problem one can separate rotations and vibrations whereas this is not possible in the classical case \cite{Iwai87c}. It would be very interesting to study the quantum reduction problem from the perspective  of reaction dynamics.
Finally we mention that shortly before the submission of our paper a related paper on the phase space structures governing reactions in rotating molecules was published \cite{KawaiKomatsuzaki11}.
That approach is based on a different normalization procedure which is treating the body angular momentum in a non-canonical manner using a rigid body approximation. It would be interesting to compare that approximation with the study in this paper in more detail.

\section*{Acknowledgments}
{\"U}.\c{C}. is grateful for the hospitality of the Johann Bernoulli Institute for Mathematics and Computer Science at the University of Groningen where the research for this paper was carried out, and to Nam\i k Kemal University for financial support during his stay.


\section*{Appendix}

\appendix

\section{The  Normal Form Algorithm}
\label{sec:NF_algorithm}

We here describe the algorithm to compute the Poincar{\'e}-Birkhoff normal form in the neighbourhood of a saddle equilibrium point of the form described in Sec.~\ref{sec:phasespacestruc_nonlinear}. The implementation of this algorithm in a  computer program is freely available \cite{Software}.

Let $H_2$ denote the quadratic Hamiltonian which gives the linearized Hamiltonian vector field at the saddle.
One says that a Hamiltonian $H$ is in normal form if
$H$ Poisson commutes with its quadratic part, i.e.
\begin{equation}
\{ H_2 , H \} := \sum_{k=1}^\dof   \big( \frac{\partial H_2}{\partial q_k} \frac{\partial H}{\partial p_k}  -  \frac{\partial H}{\partial q_k} \frac{\partial H_2}{\partial p_k} \big)=0 \,.
\end{equation}
In general $H$ is not in normal form. However, for any given order $n_0$ of the Taylor expansion of $H$ one can find a symplectic transformation to new phase space coordinates
 in terms of  which the transformed $H$ truncated at order $n_0$ is in normal form. 
This symplectic transformation is constructed from a sequence of the form 
\begin{equation}
(q_i,p_i)\equiv \vz\equiv \vz^{(0)} \mapsto \vz^{(1)} \mapsto \vz^{(2)} \mapsto \vz^{(3)}
\mapsto \ldots \mapsto \vz^{(n_0)}\,,
\label{eq:seq_trafos_z}
\end{equation}
where  $\vz^{(n)} $ is obtained from $\vz^{(n-1)} $ 
by means of a symplectic transformation 
\begin{equation}
\vz^{(n-1)} \mapsto \vz^{(n)} =  \phi_{W_n} \vz^{(n-1)} \,.
\label{eq:zn-1twozn}
\end{equation}
generated by a polynomial $W_n(\vz)$ of order $n$, i.e.
\begin{eqnarray}
W_n \in  \mathcal{W}^n := \mathrm{span} \left\{ q_1^{\alpha_1} \ldots q_\dof^{\alpha_\dof}
    p_1^{\beta_1} \ldots p_\dof^{\beta_\dof}  :   |\alpha|+|\beta|= n \right\} \, .
\label{2-09_cl}
\end{eqnarray}
Here  $|\alpha| = \sum_{k=1}^\dof \alpha_k$, $|\beta| = \sum_{k=1}^\dof \beta_k$.
More precisely, the  $\phi_{W_n}$ in \eqref{eq:zn-1twozn} denote the time-one maps of the flows generated by the Hamiltonian vector fields corresponding to the polynomials $W_n$ (see \cite{Waalkensetal08} for the details). 
The maximum order $n_0$ in \eqref{eq:seq_trafos_z} is the desired order of accuracy at which the expansion will be terminated and truncated. 

Expressing the Hamiltonian $H$ in the coordinates  $\vz^{(n)}$, $n=1,\ldots,n_0$, 
we get a sequence of Hamiltonians $H^{(n)}$,
\begin{equation}
H\equiv H^{(0)} \rightarrow H^{(1)} \rightarrow H^{(2)} \rightarrow H^{(3)}
\rightarrow \ldots \rightarrow H^{(n_0)}\,, 
\label{eq:seq_trafos_cl}
\end{equation}
where for $n=1,\ldots,n_0$,
$
H^{(n)} (\vz^{(n)}) = H^{(n-1)} (\vz^{(n-1)}  ) = H^{(n-1)} ( \phi_{W_n}^{-1} \vz^{(n)})     
$, i.e. 
\begin{equation}
H^{(n)}  = H^{(n-1)} \circ  \phi_{W_n}^{-1}\,.
\end{equation}
To avoid a proliferation of notation we will in the following neglect the superscripts $(n)$ for the phase space coordinates.

In the first transformation in \eqref{eq:seq_trafos_z} we
shift the equilibrium point
$\vz_0$ to the origin, i.e. $\vz \mapsto  \phi_{W_1} (\vz) := \vz - \vz_0$. This gives  
\begin{equation}
  H^{(1)}(\vz) = H^{(0)} (\vz+\vz_0) \, .
\label{2-02_cl}
\end{equation}
The next steps of the normal form procedure rely on  the power series
expansions of $H^{(n)}$, 
\begin{equation}
  H^{(n)}(\vz) = E_0 + \sum_{s=2}^\infty H_s^{(n)}(\vz) \, ,
\label{2-03_cl}
\end{equation}
where the $H_s^{(n)}$ are homogenous polynomials in $ \mathcal{W}^n $:
\begin{equation}
  H_s^{(n)}(\vz)  =  \displaystyle \sum_{|\alpha|+|\beta|=s}
    \frac{H_{\alpha_1,\ldots,\alpha_\dof,\beta_1,\ldots,\beta_\dof}^{(n)}}{\alpha_{1}!
      \ldots \alpha_{\dof}! \beta_{1}! \ldots \beta_{\dof}! }   
    \, q_1^{\alpha_1} \ldots q_\dof^{\alpha_\dof} p_1^{\beta_1} \ldots
    p_\dof^{\beta_\dof}  \, .
 \label{2-04_cl}
\end{equation}
For $n=1$, the coefficients in \eqref{2-04_cl} are given by the Taylor expansion of $H^{(1)}$ about the origin.
\begin{equation}
   H_{\alpha_1,\ldots,\alpha_\dof,\beta_1,\ldots,\beta_\dof}^{(1)} 
  = \displaystyle \left.
  \prod_{k,l=1}^\dof \frac{\partial^{\alpha_k}}{\partial
    q_k^{\alpha_k}} \frac{\partial^{\beta_l}}{\partial
    p_l^{\beta_l}} 
  H^{(1)}(\vz) \right|_{\vz={\bf 0}} \!\!\! .
  \label{2-04.1_cl}
\end{equation}
For $n\ge 3$, the coefficients in \eqref{2-04_cl}   are obtained recursively. For $n=2$, i.e. 
the second step in the sequence of transformations \eqref{eq:seq_trafos_z}, the coefficients in  \eqref{2-04_cl} are determined by
 a linear transformation of the phase space coordinates according to
\begin{equation}
\vz \mapsto  \phi_{W_2} (\vz) :=  M\, \vz\,.
\end{equation}
Here,  $M$ is a symplectic $2f \times 2f$ matrix which is chosen in such a way that
the transformed  Hamiltonian function
\begin{equation}
  H^{(2)}(\vz) = H^{(1)}(M^{-1} \vz)
\label{2-04.2}
\end{equation}
assumes the same form as in \eqref{eq:Hquadratic}. Section 2.3 of Ref.~\cite{Waalkensetal08} provides an explicit procedure for
constructing the transformation matrix $M$. 

For the first two steps in the sequence \eqref{eq:seq_trafos_z}, we actually did not give explicit expressions for the generating functions $W_1$ and $W_2$. For conceptual reasons (and to justify the notation) it is worth mentioning that such expression can be determined (see \cite{Waalkensetal08}).  The next  steps in  \eqref{eq:seq_trafos_z} though rely on the explicit computation of the generating functions $W_n$ with $n\ge3$.
To this end it is convenient to introduce the adjoint operator associated with a phase space function $A$:
\begin{equation}
  \mathrm{ad}_A : B \mapsto \mathrm{ad}_A B \equiv \{ A, B \}
  \, .
\label{2-08_cl}
\end{equation}
The transformation \eqref{eq:zn-1twozn} then
leads to a transformation of the Hamilton function $H^{(n-1)}$ to $H^{(n)}$ with $n \ge 3$ which in terms of the adjoint operator 
 reads
\begin{equation}
  H^{(n)} = \sum_{k=0}^\infty \frac{1}{k!} \left[ \mathrm{ad}_{W_n} \right]^k
  H^{(n-1)} \, .
\label{2-10_cl}
\end{equation}
In terms of the Taylor expansion defined in
Eqs.~(\ref{2-03_cl}-\ref{2-04.1_cl}) the transformation introduced by
Eq.~(\ref{2-10_cl}) reads
\begin{equation}
  H^{(n)}_s = \sum_{k=0}^{\left\lfloor \frac{s}{n-2} \right\rfloor}
  \frac{1}{k!} \left[ \mathrm{ad}_{W_n} \right]^k H^{(n-1)}_{s-k(n-2)} \, ,
\label{2-11_cl}
\end{equation}
where $\lfloor \cdot \rfloor$ gives the integer part of a number,
i.e., the `floor'-function. 

Using Eq.~(\ref{2-11_cl}) one finds that
the transformation defined by \eqref{2-10_cl}  satisfies the following
important properties for $n \ge 3$. Firstly, at step $n$, $n\ge3$, the terms of order less than $n$ in the power series of the Hamiltonian are unchanged, i.e.
\begin{equation}
  H_s^{(n)} = H_s^{(n-1)} \, , \;\;\; \mathrm{for} \;\;\; s < n \, ,
\label{2-12_cl}
\end{equation}
so that, in particular, $H_2^{(n)} = H_2^{(2)}$.
Defining
\begin{equation}
  \mathcal{D} \equiv \mathrm{ad}_{H_2^{(2)}} = \{ H_2^{(2)} , \cdot \} 
\label{2-14_cl}
\end{equation}
we get for the term of order $n$,
\begin{equation}
  H_n^{(n)} = H_n^{(n-1)} - \mathcal{D} W_n \, .
\label{2-13_cl}
\end{equation}
This is the so-called {\it homological  equation} which will determine the generating functions $W_n$ for $n\ge3$ from 
requiring   $\mathcal{D}
H_n^{(n)} = 0$, or equivalently $H_n^{(n)}$ to be in the kernel of the
restriction of $\mathcal{D}$ to $\mathcal{W}^n$. 
In view of \eqref{2-13_cl}  this condition yields
\begin{equation}
  H_n^{(n-1)} - \mathcal{D} W_n \in \mathrm{Ker} \,\mathcal{D} |_{\mathcal{W}^n} \, .
\label{2-15_cl}
\end{equation}
Section 3.4.1 of Ref.~\cite{Waalkensetal08} provides the explicit
procedure of finding the solution of Eq.~(\ref{2-15_cl}).  In the generic situation where the
linear frequencies $\omega_2,\ldots,\omega_\dof$ in \eqref{eq:Hquadratic} are
rationally independent, i.e. $m_2\omega_2+\ldots+m_\dof \omega_\dof =0$
implies $ m_2=\ldots=m_\dof=0$ for all integers $m_2,\ldots,m_\dof$, it
follows that for odd $n$, $H_n^{(n)} =0$, and for even $n$,
\begin{equation}
  H_n^{(n)} \in \mathrm{span} \left\{ {\cal I}^{\alpha_1} I_2^{\alpha_2} I_3^{\alpha_3}
    \ldots I_\dof^{\alpha_\dof}  : |\alpha|=n/2 \right\} \, ,
\label{2-15.2}
\end{equation}
where ${\cal I} = (p^2_1- q^2_1)/2$ and $I_k = (q_k^2 + p_k^2)/2$, with $k = 2, \ldots, \dof$.

Applying the transformation (\ref{2-10_cl}), with the generating function
defined by \eqref{2-13_cl}, for $n=3,\ldots,n_0$, and truncating the
resulting power series at order $n_0$ one
arrives at the Hamiltonian $H_{\mathrm{NF}}^{(n_0)}$ corresponding to
the $n_0^{\mathrm{th}}$ order {\it normal form} (NF) of the
Hamiltonian $H$:
\begin{equation}
  H_{\mathrm{NF}}^{(n_0)}(\vz) = E_0 + \sum_{s=2}^{n_0} H_s^{(n_0)}(\vz) \, .
\label{2-15.5_cl}
\end{equation}
We stress that $H_{\mathrm{NF}}^{(n_0)}$ represents an
$n_0^\mathrm{th}$ order approximation of the original Hamiltonian $H$ obtained from
expressing $H$ in terms of the \emph{normal form coordinates} $\vz_{\mathrm{NF}} $ which in turn are obtained from the symplectic transformation of the original coordinates $\vz=(q_i,p_i)$ 
\begin{equation}
\vz_{\mathrm{NF}} = \phi (\vz) =  (\phi_{W_{n_0}} \circ \phi_{W_{n_0-1}}    \circ   \cdots  \circ \phi_{W_{2}}    \circ  \phi_{W_{1}})(\vz)  \,.
\label{eq:def_U}
\end{equation} 
This  is why it is legitimate to use
$H^{(n_0)}_{\mathrm{NF}}$ instead of $H$ in analyzing the dynamics in the neighbourhood of the saddle.

We emphasize that the full procedure to compute $H_{\mathrm{NF}}^{(n_0)} $ and the corresponding coordinate transformation is algebraic
in nature, and can be implemented on a computer. A computer program is freely available from \cite{Software}.



\end{document}